\documentstyle[twocolumn,prl,aps,epsfig,floats]{revtex}

\newcommand{\PLB}[3]{{\it Phys.~Lett.}~{\bf B#1}, #2~(19#3)}

\newcommand{\PRD}[3]{{\it Phys.~Rev.}~{\bf D#1}, #2~(19#3)}

\newcommand{\gtsim}{\stackrel{\scriptscriptstyle>}{\scriptscriptstyle\sim}}

\def\D0{D\O~}

\begin{document}
\draft

\twocolumn[\hsize\textwidth\columnwidth\hsize\csname
@twocolumnfalse\endcsname

\title{
Higgs Boson Production at the LHC with Soft Gluon Effects
}

\author{
C.~Bal\'azs$^{1,2}$ ~ and ~ C.--P. Yuan$^{3,4}$
}
\address{
$^1$~Department of Physics and Astronomy, 
     University of Hawaii, Honolulu, HI 96822, U.S.A. \\
$^2$~Fermi National Accelerator Laboratory, Batavia, Illinois 60510, U.S.A. \\
$^3$~National Center for Theoretical Sciences,
         National Tsing-Hua University, Taiwan, Republic of China \\
$^4$~Department of Physics and Astronomy, Michigan State University,
         East Lansing, MI 48824, U.S.A.
}

\date{January 12, 2000}

\maketitle
\thispagestyle{empty}

\begin{abstract}

We present results of QCD corrections to Higgs boson production at the 
CERN Large Hadron Collider. Potentially large logarithmic contributions 
from multiple soft--gluon emission are resummed up to all order in the 
strong coupling. Various kinematical distributions, including the Higgs 
transverse momentum, are predicted at the ${\cal O}(\alpha_s^3)$ level. 
Comparison is made to outputs of the popular Monte Carlo event generator 
PYTHIA.

\end{abstract}


\vskip2pc]


\section{Introduction}


One of the fundamental questions of the Standard Model (SM) of elementary 
particle physics is the dynamics of the electroweak symmetry 
breaking. Within the SM, the Higgs mechanism postulates the existence of a 
scalar field, the elementary excitation of which is called the Higgs 
boson. Four experimental collaborations at the LEP II collider search for 
the Higgs boson in the $e^+ e^- \to Z^0 H$ process up to 202 GeV center of 
mass energy. DELPHI and L3 set a preliminary exclusion limit of $m_H > 96$ 
GeV on the Higgs mass, followed closely by the limit of $m_H > 91$ GeV set 
by ALEPH and OPAL \cite{Felcini:1999yh}. According to recent preliminary 
information, the combined lower limit is close to $m_H \gtsim 106$ GeV 
\cite{LEPC}. Global fits to electroweak observables appear to prefer a low 
mass Higgs particle, with a mean value close to 100 GeV, and less than 250 
GeV within 95 percent of confidence \cite{EWGlobalFits,mWmt}. 



Among other aims the main goal of the next proton-proton accelerator, the 
14 TeV center of mass energy Large Hadron Collider (LHC) at CERN, is to 
establish the existence of the Higgs boson and to measure its basic 
properties. At the LHC a light SM Higgs boson will be mainly produced 
through the partonic subprocess $g g$ (via top quark loop) $\to H X$ 
\cite{Spira:1997zs}. It can be detected, after about 1.5 years of running 
with a statistical significance of at least 4, in its $H \to \gamma 
\gamma$ decay mode, if its mass is in the 100-150 GeV range 
\cite{AtlasCPandCMSTDR}. If the Higgs mass is higher than about 150 GeV, 
then its $H \to Z^{0(*)} Z^0$ decay mode is the cleanest and most 
significant \cite{AtlasCPandCMSTDR}. In this letter our focus is on the 
Higgs boson production, and in our numerical illustration we choose $m_H = 
150$ GeV.


According to earlier studies, a statistical significance on the order of 5%
-10 can be reached for the inclusive $H \to \gamma \gamma$ and for the $H 
+ \mbox{jet} \to \gamma \gamma + \mbox{jet}$ signals, although actual 
values depend on luminosity and background estimates. In 
Ref.~\cite{Abdullin} it was found that in order to optimize the 
significance it is necessary to impose a 30 GeV cut on the transverse 
momentum of the jet, or equivalently (at next-to-leading order precision), 
on the transverse momentum ($Q_T$) of the photon pair. With this cut in 
place, extraction of the signal in the Higgs + jet mode requires the 
precise knowledge of both the signal and background distributions in the 
medium to high $Q_T$ region. 


To reliably predict the $Q_T$ distribution of Higgs bosons at the LHC, 
especially for the low to medium $Q_T$ region where the bulk of the rate is, 
the effects of the multiple soft--gluon emission have to be included. One 
approach to achieve this is parton showering \cite{Sjostrand:1985xi}. 
Although the universality of this method makes it a very powerful tool, 
present drawbacks of this ansatz are the lack of the proper normalization 
(which takes into account the full fixed order QCD corrections), the lack 
of exact matrix elements even in the high $Q_T$ region, and the lack of 
uniqueness of the prediction ("tunability"). There is ongoing work 
to correct these problems \cite{StirlingSeymourSjostrandMrenna}. 


A more reliable prediction of the Higgs $Q_T$ can be obtained utilizing the 
Collins-Soper-Sterman (CSS) resummation formalism \cite{CS,CSS}, which takes 
into account the effects of the multiple soft--gluon emission while 
reproducing the rate, systematically including the higher order 
corrections. It is possible to smoothly match the CSS result to the fixed 
order one in the medium to high $Q_T$ region, thus obtaining the best prediction 
in the full $Q_T$ region \cite{BalazsYuanWZ}. Compared to fixed orders, the 
resummed result depends on a few extra parameters. These parameters are 
new renormalization scales ($C_i$, only two of which are independent) 
\cite{CSS}, and a few universal, non-perturbative parameters ($g_i$), 
which are extracted from present experiments and then used to predict the 
results of future ones \cite{BrockLadinskyLandryYuan}. In this letter, we 
use this formalism to calculate the total cross sections and $Q_T$ 
distributions of Higgs bosons at the LHC. 


Our results here, together with the resummed calculations for the 
diphoton and $Z^0$ boson pair production 
\cite{BalazsBergerMrennaYuan,BalazsYuanZZ,BalazsNadolskySchmidtYuan,Thesis}, 
provide a consistent set of QCD calculations of the transverse momentum 
(and other) distribution(s) of the Higgs bosons and their backgrounds at 
the LHC. These results systematically include both the multiple soft--%
gluon effects and the finite order QCD corrections, and can be used to 
tune the shower MC's which experimentalists extensively use when 
extracting the Higgs signal, or can be utilized independently by means 
of the ResBos Monte Carlo event generator \cite{BalazsYuanWZ}.

\section{Analytical Results}


Within the SM \cite{Spira:1997zs,Georgi:1978}, as well as in the minimal 
supersymmetric standard  model (MSSM) with small $\tan\beta$ 
\cite{Spira:1997zs}, the dominant production mode of neutral Higgs bosons 
at the LHC is gluon fusion via a heavy quark loop. The lowest order cross 
section of this process is formally ${\cal O}(\alpha_s^2)$ in the strong 
coupling. Fixed order QCD corrections to this production mechanism are 
known to substantially increase the rate. For a light Higgs boson the 
${\cal O}(\alpha_s^3)$ to ${\cal O}(\alpha_s^2)$ $K$-factor is in the 
order of 2 (cf. Fig. \ref{Fig:Tot}). The full ${\cal O}(\alpha_s^4)$ 
calculation is not completed yet, but the real emission 
\cite{KauffmanDesaiRisal} and the virtual contributions \cite{Schmidt} are 
separately available. 
%
%
Resummed calculations, taking into account the soft--gluon effect, were also 
performed to estimate the size of the uncalculated higher order corrections 
\cite{Kramer:1996iq}, as well as to predict the shape of the Higgs transverse 
momentum distribution \cite{HinchliffeKauffmanKao,Yuan}. 


In this work we use the Collins-Soper-Sterman (CSS) soft--gluon resummation
formalism to calculate the QCD corrections from the multiple--soft gluon 
emission. Calculations similar to this were earlier performed in 
Refs. \cite{HinchliffeKauffmanKao,Yuan}. Our present calculation improves 
these by including ${\cal O}(\alpha_s^4)$ terms in the Sudakov exponent, 
by applying the state of the art matching to the latest fixed order 
distributions, by using a QCD improved gluon-Higgs effective coupling 
\cite{Kniehl:1995tn}, by utilizing an improved non-perturbative function, 
and by including 
the effect of the Higgs width. 

We also utilize the approximation that the object
which couples the gluons to the Higgs (the top quark in the SM) is much
heavier than the Higgs itself. This approximation is not essential to our
calculation and can be released by including the complete Wilson 
coefficients with all the relevant masses. The heavy quark approximation in
the SM has been shown to be reliable within 5 percent for $m_H < 2 m_t$
\cite{Graudenz:1993,Kunszt:1996yp,FlorianGrazziniKunsztH}, and still
reasonable even in the range of $m_H \gtsim 2 m_t$ \cite{Kramer:1996iq}. It
has also been shown that the approximation remains valid for the $Q_T$
distribution in the large $Q_T$ region, provided that $m_H < m_t$ and $Q_T <
m_t$ \cite{Baur:1990cm}. In this work we assume that the approximation is
valid in the whole $Q_T$ region. In the MSSM the heavy quark approximation
is also a reliable ansatz for the case of the light Higgs boson and small
$\tan\beta$ when the Yukawa coupling of the bottom quark is negligible
(c.f. \cite{Kramer:1996iq} and references therein). 
%
%
%
Using the CSS formalism we resum large logs of the type $\ln(Q/Q_T)$ in the low 
$Q_T$ region, and we match the resummed result to the fixed order calculation 
which is valid for high $Q_T$ \cite{BalazsYuanWZ}. We also include the $qg$ and 
$q\bar{q}$ subprocesses which, in combination, can constitute up to 10 percent 
of the total rate, depending on the Higgs mass \cite{Graudenz:1993}.

The resummed differential cross section of a neutral Higgs boson, denoted by
$\phi^0$ in the SM or MSSM, produced in hadronic collisions is written as
\begin{eqnarray}
&&{\frac{d\sigma (h_1h_2\to \phi^0X)}{dQ^2\,dQ_T^2\,dy}} = 
\sigma_0\, \frac {Q^2}{S}\, 
{Q^2\Gamma_\phi/m_\phi \over (Q^2-m^2_\phi)^2+(Q^2\Gamma_\phi/m_\phi)^2}
   \nonumber \\ && \times
\left\{ {\frac 1{(2\pi )^2}}\int d^2b\,e^{i{\vec{Q}_T}\cdot
{\vec{b}}} 
{\widetilde{W}_{gg} (b_{*},Q,x_1,x_2,C_{1,2,3})}
   \right. \nonumber \\ && ~~~~\times \left.
\widetilde{W}_{gg}^{NP}(b,Q,x_1,x_2) 
+ Y(Q_T,Q,x_1,x_2,C_4) \right\} .
\label{Eq:ResFor}
\end{eqnarray}
The kinematical variables $Q$, $Q_T$, and $y$ are the invariant mass,
transverse momentum, and rapidity of the Higgs boson, respectively, 
in the laboratory frame. The parton momentum fractions are defined as 
$x_1=e^{y}M_T/\sqrt{S}$, and $x_2=e^{-y}M_T/ \sqrt{S}$, with 
$M_T = \sqrt{Q^2+Q_T^2}$, and $\sqrt{S}$ being the center--of--mass
(CM) energy of the hadrons $h_1$ and $h_2$. The lowest order cross
section is
\begin{equation}
\sigma_0 = \kappa_\phi(Q) {\sqrt{2}G_F\alpha^2_s(Q^2) \over 576 \pi},
\end{equation}
where $G_F$ is the Fermi constant, and $\kappa_\phi$, the QCD corrected
effective coupling of the Higgs boson to gluons in the heavy top quark limit
(cf. Ref.\cite{Kramer:1996iq}), is defined as
\begin{eqnarray}
\kappa_\phi(Q) & = & 1 + \frac{11}{2} \frac{\alpha^{(5)}_s(m_t^2)}{\pi}
+ \frac{3866 - 201\, N_f}{144}\left( \frac{\alpha^{(5)}_s(m_t^2)}{\pi}\right)^2
\nonumber \\
& & + \frac{153-19\, N_f}{33-2\, N_f}
~\frac{\alpha^{(5)}_s(Q^2) - \alpha^{(5)}_s(m_t^2)}{\pi}
+ {\cal O}(\alpha_s^3) 
\end{eqnarray}
where $\alpha^{(5)}_s$ is the strong coupling constant in the 
$\overline{\rm MS}$ scheme with 5 active flavors, and $m_t$ denotes the 
pole mass of the top quark. 

The renormalization group invariant kernel of the Fourier integral 
$\widetilde{W}_{gg}(b_{*},Q,x_1,x_2,C_{1,2,3})$, and the $Q_T$ regular term 
$Y(Q_T,Q,x_1,x_2,C_4)$, 
together with the variables $b_*$ and $C_1$ to $C_4$, are given in Ref. 
\cite{Yuan}.
The definition of $\widetilde{W}_{gg}$, contains the Sudakov exponent
\begin{eqnarray}
&& {\mathcal S}(Q,b_*,C_1,C_2) =
\nonumber \\ 
&& ~~ \int_{C_1^2/b_*^2}^{C_2^2 Q^2}\frac{d\overline{\mu }^2}{%
\overline{\mu }^2}\left[ 
A\left( \alpha_s(\overline{\mu }),C_1\right) \ln
\left( \frac{C_2Q^2}{\overline{\mu }^2}\right) + 
\right. \nonumber \\ 
&& \left. ~~~~~~~~~~~~~~~~~~~~ B\left( \alpha_s(\overline{\mu}),C_1,C_2\right) 
\right].
\end{eqnarray}
In the perturbative expansion of the $A\left( \alpha_s(\overline{\mu }),C_1\right)$
and $B\left( \alpha_s(\overline{\mu}),C_1,C_2\right)$ functions we follow
the notation of Ref. \cite{ArnoldKauffman}. 
In our present calculation, we include the 
process independent next-to-next-to-leading order coefficient
\begin{eqnarray}
&& A^{(2)}(C_1) = \nonumber \\ 
&& ~~ 4 C_A\left[ \left( {\frac{67}{36}}-{\frac{\pi ^2}{12}}\right)
N_C-{\frac 5{18}}N_f-2 \beta _1\ln \left( {\frac{b_0}{C_1}}\right) \right] ,
\end{eqnarray}
in the expansion of the $A\left( \alpha_s(\overline{\mu }),C_1\right)$ 
function, where $C_A = 3$ is the Casimir of the 
adjoint representation of $SU(3)$, $N_C = 3$ is the number of colors, and $N_f = 
5$ is the number of active quark flavors. With the inclusion of $A^{(2)}$ the 
only missing next-to-next-to-leading order contribution in the Sudakov exponent 
is the $B^{(2)}$ term, which is suppressed by 
$1/\ln \left( \frac{Q^2}{Q_T^2}\right)$ 
with respect to $A^{(2)}$, 
and by $\alpha_s$ with respect to $B^{(1)}$.
This is illustrated by the expansion of the asymptotic part of the 
cross section:
\begin{eqnarray}
&& \lim_{Q_T\rightarrow 0}\frac{d\sigma }{dQ^2dQ_T^2dy}= \nonumber \\
&& \sigma _0\frac
1{Q_T^2}\sum_{i,j}\sum_{n=1}^{\infty}\sum_{m=0}^{2n-1}\left( 
\frac{\alpha _s(Q)}\pi \right)^n
C_{nm}^{(ij)}\ln^m \left( \frac{Q^2}{Q_T^2}\right) ,
\end{eqnarray}
where $i$ and $j$ label incoming partons. 
While the $A^{(2)}$ coefficient contributes to the above series via
\begin{eqnarray}
&& C_{21}^{(ij)} \propto \nonumber \\
&& -\frac 12\left[ \left( B^{(1)}\right) ^2-A^{(2)}-\beta _0A^{(1)}\ln
\left( \frac{\mu _R^2}{Q^2}\right) -\beta _0B^{(1)}\right] f_i\, f_j,
\nonumber \\
\end{eqnarray}
the $B^{(2)}$ coefficient only occurs 
as
\begin{eqnarray}
&& C_{20}^{(ij)} \propto \nonumber \\
&& \left[ \zeta (3)\left( A^{(1)}\right) ^2+\frac 12B^{(2)}+\frac
12\beta _0B^{(1)}\ln \left( \frac{\mu _R^2}{Q^2}\right) \right] f_i\,
f_j,
\end{eqnarray}
where our notation coincides with that of Ref. \cite{ArnoldKauffman}.
Hence, the contribution from the uncalculated $B^{(2)}$, compared to that 
from $A^{(2)}$, is expected to be smaller because of the additional log 
weighting the $A^{(2)}$ coefficient.
To estimate the size of the contribution from $B^{(2)}$, we follow the usual 
practice in a perturbative calculation by varying the renormalization constants 
($C_1$ and $C_2$) in the Sudakov factor by a factor of 2. 
The results are shown in Fig. \ref{Fig:PYTHIA}. 

The form of our non-perturbative function $\widetilde{W}_{gg}^{NP}$
coincides with the one used for the $gg \to \gamma \gamma$ process in Ref.
\cite{BalazsBergerMrennaYuan} 
\begin{eqnarray} &&
\widetilde{W}_{gg}^{NP}(b,Q,Q_0,x_1,x_2) = 
\nonumber \\ &&
{\rm exp}\left[ -g_1b^2-{\frac{C_A}{C_F}}g_2b^2\ln
\left( {\frac Q{2Q_0}}\right) -g_1g_3b\ln {(100x_1x_2)} \right],
\nonumber \\
\label{Eq:WNP} \end{eqnarray}
where the Casimir of the fundamental $SU(3)$ representation is $C_F = 4/3$. 
The values of the non-perturbative
parameters $g_i$ are defined in Ref.~\cite{BalazsNadolskySchmidtYuan}.
The uncertainties of the resummed distribution, stemming from the 
non-perturbative function, were found to be in the order of 5 percent 
(cf. \cite{BalazsYuanLongH}).
In the high $Q_T$ region Eq. (\ref{Eq:ResFor}) is matched to the fixed order
perturbative result (at the ${\cal O}(\alpha_s)$) of 
Ref. \cite{FlorianGrazziniKunsztH} in the manner described in 
Ref. ~\cite{BalazsYuanWZ}.

\section{Numerical Results}

The analytic results are coded in the ResBos Monte Carlo event generator
\cite{BalazsYuanWZ,Thesis}, which uses the following electroweak input
parameters \cite{PDB}: 
\begin{eqnarray} &&
G_F = 1.16639\times 10^{-5}~{\rm GeV}^{-2}, ~~ m_Z = 91.187~{\rm GeV}, 
\nonumber \\ && 
m_W = 80.36~{\rm GeV}.
\end{eqnarray} 
As in the background calculation \cite{BalazsYuanZZ}, 
we use the canonical choice of the
renormalization constants ($C_1=C_3=2e^{-\gamma _E}\equiv C_0$ and $%
C_2=C_4=1$ \cite{CSS}), the NLO expressions for the running
electromagnetic and strong couplings $\alpha(\mu)$ and $\alpha_S(\mu)$, as
well as the NLO parton distribution function set CTEQ4M (defined in the
modified minimal subtraction, $\overline{MS}$, scheme) \cite{CTEQ4}. We set the
renormalization scale equal to the factorization scale: $\mu_R=\mu_F=Q$. In
the choice of the non-perturbative parameters, we follow
Ref.~\cite{BalazsNadolskySchmidtYuan}. Since we are not concerned with the
decays of Higgs bosons in this work, we do not impose any kinematic cuts.
We defer the more extensive study, including various decay modes and QCD
backgrounds, to a future publication Ref. \cite{BalazsYuanLongH}.

\begin{figure}[t]
\epsfig{file=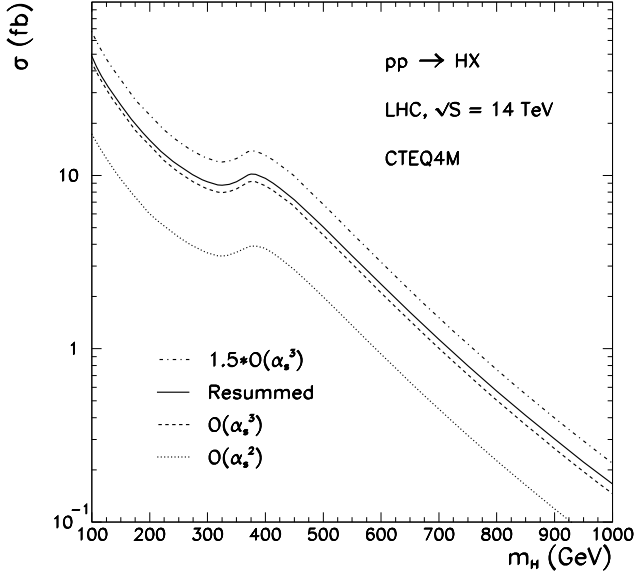,width=8.65cm}
\caption[Fig:Tot]{
SM Higgs boson production cross sections at the LHC via gluon fusion as the 
function of the Higgs mass, with QCD corrections calculated by soft--gluon 
resummation (solid), at fixed order ${\cal O}(\alpha_s^3)$ (dashed), and without 
QCD corrections at ${\cal O}(\alpha_s^2)$ (dotted). The ${\cal O}(\alpha_s^3)$ 
curve is scaled by 1.5 (dash-dotted, c.f. Ref. \cite{Kramer:1996iq}) to 
estimate the ${\cal O}(\alpha_s^4)$ result.
}
\label{Fig:Tot}
\end{figure}

Fig.~\ref{Fig:Tot} displays Higgs boson production cross sections via the 
gluon fusion process at the LHC, calculated with various QCD corrections
in the SM as the function of the Higgs mass. 
The ratio of the fixed order ${\cal O}(\alpha_s^3)$
(dashed) and the lowest order ${\cal O}(\alpha_s^2)$ (dotted) curves varies
between 2.0 and 2.3. We note that less than 2 percent of the ${\cal
O}(\alpha_s^3)$ corrections come from the $qg$ and $q {\bar q}$ initial
states for Higgs masses below 200 GeV. The resummed curve is about
10 percent higher than the ${\cal O}(\alpha_s^3)$ one, as expected based on
the findings that the CSS formalism preserves the fixed order rate
within the error of the matching (the latter being higher order)
\cite{BalazsYuanWZ}. The resummed rate is close to the ${\cal
O}(\alpha_s^3)$, because we used the ${\cal O}(\alpha_s^3)$ fixed order
results to derive the Wilson coefficients which are utilized in our
calculation. In Ref. \cite{Kramer:1996iq} a resummed calculation estimated the 
size of the ${\cal O}(\alpha_s^4)$ corrections, and a typical value of 1.5 of 
the ${\cal O}(\alpha_s^4)$ to ${\cal O}(\alpha_s^3)$ $K$-factor can be inferred 
from that work. Based on this, we also plot the ${\cal O}(\alpha_s^3)$ curve 
rescaled by 1.5, to illustrate the possible size of the ${\cal O}(\alpha_s^4)$ 
corrections and to establish the normalization of our resummed calculation 
among the fixed order results.

\begin{figure}[t]
\epsfig{file=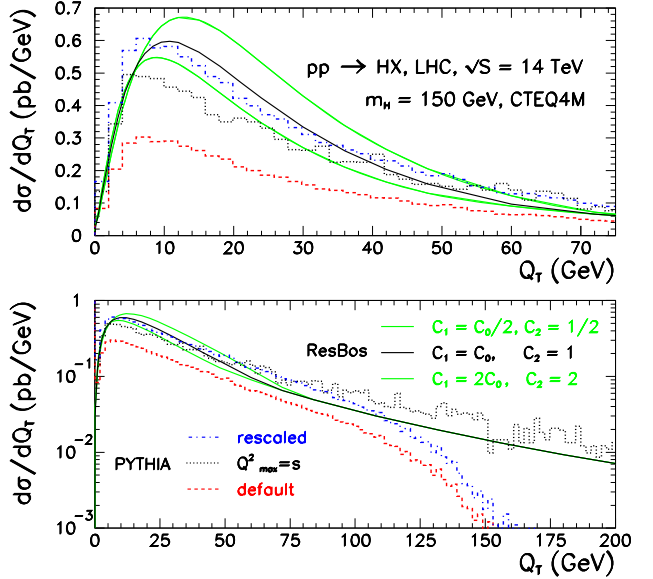,width=8.65cm}
\vskip1pc
\caption[Fig:PYTHIA]{
Higgs boson transverse momentum distributions calculated by ResBos (curves) 
and PYTHIA (histograms). The default (middle) ResBos curve was calculated with 
the canonical choice of the renormalization constants, and the other two 
with doubled (lower curve) and halved (upper curve) values of $C_1$ and 
$C_2$. For PYTHIA we show the original output with default input 
parameters (dashed), the same rescaled by a factor of $K = 2$ (dotted), 
and a curve calculated by the altered input parameter value $Q_{max}^2 = 
s$ (dash-dotted). The lower portion, with a logarithmic scale, also shows 
the high $Q_T$ region.
}
\label{Fig:PYTHIA}
\end{figure}

Fig.~\ref{Fig:PYTHIA} compares the Higgs boson transverse momentum distributions
calculated by ResBos (curves) and by PYTHIA \cite{PYTHIA} (histograms from 
version 6.122). The middle solid curve is calculated using the canonical choice 
for the renormalization constants in the Sudakov exponent: $C_1 = C_0$, and 
$C_2 = 1$. To estimate 
the size of the uncalculated $B^{(2)}$ term, we varied these 
renormalization constants multiplying both by 1/2 and 2. The upper solid 
curve shows the result for $C_1 = C_0/2, C_2 = 1/2$, and the lower solid 
curve for $C_1 = 2C_0, C_2 = 2$. The band between these two curves gives 
the order of the uncertainty following from the exclusion of $B^{(2)}$.
The typical size of this uncertainty, e.g. around the peak region, is in the
order of $\pm 10$ percent. The corresponding uncertainty in the total cross 
section is also in the same order. The dashed PYTHIA histogram is plotted without
altering its output. The normalization of PYTHIA, as that of any parton
shower MCs, is the lowest order ${\cal O}(\alpha_s^2)$ for this process. 
The default PYTHIA histogram is also plotted after the rate is multiplied by
the factor $K = 2$ (dotted). The shape of the PYTHIA histogram agrees reasonably
with the resummed curve in the low and intermediate $Q_T$ ($\lesssim
125$ GeV) region. For large $Q_T$ the PYTHIA prediction falls under the
ResBos curve, since ResBos mostly uses the exact fixed order ${\cal
O}(\alpha_s^3)$ matrix elements in that region (c.f. Ref.
\cite{BalazsYuanWZ}), while PYTHIA still relies on the multi-parton
radiation ansatz. PYTHIA can be tuned to agree with ResBos in the high $Q_T$ 
region (dash-dotted), by changing the maximal virtuality a parton can acquire
in the course of the shower, i.e. the $Q_{max}^2$ parameter, from the default 
value to the partonic center of mass energy $s$. In that case, however, the low 
$Q_T$ region will have disagreement, since the normalization in PYTHIA is 
conserved, so events in the low $Q_T$ region are depleted.

\section{Conclusions}

In this letter we presented Higgs boson production rates and $Q_T$ 
distributions for the LHC, including ${\cal O}(\alpha_s^3)$ fixed order 
QCD and multiple soft--gluonic corrections by means of the CSS 
resummation formalism. We showed that the resummed rate recovers the fixed 
order ${\cal O}(\alpha_s^3)$ rate, as expected within the CSS formalism. 
We investigated the uncertainty of the resummed prediction due to 
uncalculated terms in the Sudakov exponent. We found that the shape of the 
resummed prediction in the low $Q_T$ region is in reasonable agreement 
with the default result of PYTHIA.

\section*{Acknowledgments}

We thank the CTEQ Collaboration, and M. Spira for invaluable discussions. 
We are indebted to I. Puljak for providing us with the PYTHIA results. C. B. 
thanks the organizers of the les Houches SUSY/Higgs/QCD workshop for their 
hospitality and financial support, and the Fermilab Theory Group for 
their invitation and financial support. This work was supported in part by 
the NSF under grant PHY--9802564 and by DOE under grant DE-FG-03-%
94ER40833.




\end{document}